# ERLC (Twin LC) and LHC/FCC Based eA Colliders

A. N. Akay[a], B. Dagli[a], B. Ketenoglu[b,*], S. Sultansoy[a,c]

[a]*TOBB University of Economics and Technology, Ankara, Turkey*
[b]*Department of Engineering Physics, Ankara University, Ankara, Turkey*
[c]*ANAS Institute of Physics, Baku, Azerbaijan*
[*]*Correspondence: bketen@eng.ankara.edu.tr*

**Abstract**

Construction of the ERLC (twin LC) collider tangential to LHC or FCC will give opportunity to realize *eA* collisions at multi-TeV center-of-mass energies. Luminosity estimations show that values comparable with that of ERL60 based *eA* colliders are achievable while center of mass energies are essentially higher. Certainly, proposed *eA* colliders have great potential for clarifying QCD basics and nuclear structure.

## 1. Introduction

Recently, V. I. Telnov has proposed ERLC (twin LC) scheme to improve ILC luminosity by two orders [1]. In previous paper we have considered possible use of the ERLC for HL-LHC, HE-LHC and FCC based *ep* colliders [2]. It was shown that using ERLC is much more advantageous comparing to ERL60, which is considered as baseline option for HL/HE-LHC and FCC based lepton-hadron colliders [3].

In this paper, possible use of the ERLC for HL/HE-LHC and FCC based *eA* colliders have been considered. (for earlier *eA* collider proposals see review [4] and references therein, as well as [5] and [6] for FCC and SppC, respectively) Main parameters of ERLC electron and LHC/FCC lead beams are presented in Section 2. In next section we evaluate parameters of corresponding *eA* colliders. In Section 4, we briefly discuss physics search potential of these machines. Our conclusion and recommendations are given in Section 5.

## 2. Parameters of ERLC, LHC and FCC

In this section, we present parameters of ERLC, HL-LHC, HE-LHC and FCC, which are used for estimation of main parameters of *eA* colliders in the following section. Parameters of ERLC and ILC are given in Table 1 (Table 2 of Reference [1]). Table 2 presents parameters of HL-LHC, HE-LHC and FCC lead beams upgraded for ERL60 *eA* colliders (Table 9 in Reference [3]).

Table 1. Main parameters of ERLC and ILC

| Parameter [unit] | ERLC | ILC |
|---|---|---|
| Beam Energy [GeV] | 125 | 125 |
| N per bunch [$10^{10}$] | 0.5 | 2.0 |
| Norm. emit., $\epsilon_{x,n}$ [µm] | 20 | 10 |
| Norm. emit., $\epsilon_{y,n}$ [µm] | 0.035 | 0.035 |
| $\beta_x$ at IP [cm] | 25 | 1.3 |
| $\beta_y$ at IP [cm] | 0.03 | 0.04 |
| $\sigma_x$ at IP [µm] | 4.5 | 0.73 |
| $\sigma_y$ at IP [nm] | 6.1 | 7.7 |
| Rep. rate, $f$ [Hz] | $2 \times 10^8$ | 6560 |
| Bunch distance [m] | 1.5 | 166 |
| Duty cycle | 1/3 | n/a |

Table 2. LHC/FCC lead beam parameters upgraded for ERL60 based *eA* colliders

| Parameter [unit] | HL-LHC | HE-LHC | FCC |
|---|---|---|---|
| $E_{Pb}$ [PeV] | 0.574 | 1.03 (1.11) | 4.1 |
| $E_e$ [GeV] | 60 | 60 | 60 |
| $\sqrt{S}_{eN}$ [TeV] | 0.8 | 1.1 | 2.2 |
| Bunch spacing [ns] | 50 | 50 | 100 |
| Number of Bunches | 1200 | 1200 | 2072 |
| Ions per Bunch [$10^8$] | 1.8 | 1.8 | 1.8 |
| Norm. emit., $\epsilon_A$ [µm] | 1.5 | 1.0 | 0.9 |
| Electrons per bunch [$10^9$] | 4.67 | 6.2 | 12.5 |
| Electron current [mA] | 15 | 20 | 20 |
| IP beta function, $\beta^*_A$ [cm] | 7 | 10 | 15 |
| Hourglass factor $H_{geom}$ | 0.9 | 0.9 | 0.9 |
| Pinch Factor $H_{b-b}$ | 1.3 | 1.3 | 1.3 |
| Bunch filling $H_{coll}$ | 0.8 | 0.8 | 0.8 |
| $L_{eN}$ [$10^{32}$ cm$^{-2}$s$^{-1}$] | 7 | 18 | 54 |
| $\sqrt{S}_{eA}$ [TeV] | 3.7 | 5.1 | 9.9 |
| $L_{eA}$ (AloHEP) [$10^{30}$ cm$^{-2}$s$^{-1}$] | 2.54 | 6.35 | 17.4 |

Last two rows are added by us. Then, energy value of lead beam for HE-LHC option is incorrect, real value is 1.11 PeV instead of 1.03 PeV.

## 3. Luminosity of *eA* Colliders

In this section luminosity, disruption and beam-beam tune shift parameters have been calculated using parameters of electron and nucleus beams given in previous section. Several years ago, the software AloHEP has been developed for estimation of main parameters of linac-ring type *ep* colliders [7, 8]. Recently, AloHEP has been upgraded [9] for all types of colliders (linear, circular and linac-ring) as well as colliding beams (electron, positron, muon, proton and nuclei).

Main parameters of HL-LHC, HE-LHC and FCC based *eA* colliders, calculated by using the AloHEP software and parameters of electron (lead) beam from Table 1 (Table 2), are presented in Table 3. It is seen that luminosity of ERLC based colliders is more than 2 orders higher comparing to luminosity of ILC based ones.

Table 3. *eA* collider parameters using nominal parameters from Tables 1 and 2

| Lead beam | *e*-beam | $L_{eA}$ [cm$^{-2}$s$^{-1}$] | $\xi$ [10$^{-4}$] | D |
|---|---|---|---|---|
| HL-LHC | ILC | 5.26x10$^{27}$ | 5.3x10$^{-2}$ | 0.4 |
| | ERLC | 9.03x10$^{29}$ | 1.3x10$^{-2}$ | 0.4 |
| HE-LHC | ILC | 1.00x10$^{28}$ | 8.0x10$^{-2}$ | 0.7 |
| | ERLC | 1.70x10$^{30}$ | 2.0x10$^{-2}$ | 0.7 |
| FCC | ILC | 2.96x10$^{28}$ | 8.9x10$^{-2}$ | 2.1 |
| | ERLC | 2.31x10$^{30}$ | 2.2x10$^{-2}$ | 2.1 |

Bunch distance of LHC/FCC lead beam (bunch spacing 50 ns corresponds to bunch distance 15 m) is 10 times higher than electron beam bunch distance (1.5 m). This means that only 1/10 of electron bunches take part in *eA* collisions. Therefore, one can consider following upgrade of ERLC parameters: 10 times lower repetition rate and 10 times higher number of electrons per bunch. With this upgrade, we have obtained parameters of *eA* collisions presented in Table 4. Let us mention that $\xi$ values should be decreased by an order which can be handled by using of crab-waist scheme in collision region [10].

Table 4. *eA* collider parameters for upgraded ERLC

| Parameter [unit] | HL-LHC | HE-LHC | FCC |
|---|---|---|---|
| $\sqrt{s}_{eA}$ [TeV] | 16.9 | 23.5 | 45.3 |
| $L_{eA}$ [cm$^{-2}$s$^{-1}$] | 9.04x10$^{30}$ | 1.70x10$^{31}$ | 2.32x10$^{31}$ |
| $\sigma_{x,y}$ at IP [µm] | 5.94x10$^{-6}$ | 4.33x10$^{-6}$ | 2.52x10$^{-6}$ |
| Disruption, D | 0.4 | 0.7 | 2.1 |
| Tune Shift, $\xi$ | 1.3x10$^{-1}$ | 2.0x10$^{-1}$ | 2.2x10$^{-1}$ |

Finally, luminosity of the ERLC based *eA* colliders can be further improved using dynamic focusing scheme [11]. After all, $L_{eA} \approx 10^{32}$ cm$^{-2}$s$^{-1}$ seems achievable for all options with reasonable modifications of ERLC and LHC/FCC parameters.

## 4. Physics Potential

Energy frontier *eA* colliders are a must to providing precision PDFs for adequate interpretation of LHC/FCC *pA* and *AA* experimental data. On the other hand, construction of lepton-nucleus colliders is very important for clarifying QCD basics and nuclear structure. Physics search program of *eA* colliders were widely discussed within LHeC [12] and EIC [13] projects. Several examples are following: physics of non-linear color fields and gluon saturation, particle propagation through matter and transport properties of nuclei, parton fragmentation, production mechanism for quarkonia, confiment mechanism and so on.

Obviously, ERLC is more advantageous than ERL60 for both luminosity and center-of-mass energy aspects. Let us emphasize that ERLC will provide opportunity to increase the electron beam energy by extending the length of linacs. For instance, 1 TeV $e^+e^-$ will enhance the $\sqrt{S_{eA}}$ by a factor of two, which allows investigation of 4-times smaller values of *x* Bjorken (one order smaller x Bjorken comparing to ERL60 based *eA* colliders).

## 5. Conclusion

Certainly, the ERLC [1] opens new horizons for linear e$^+$e$^-$ colliders. It seems that the ERLC may have essentially impact on future linac-ring type lepton-hadron colliders, too. The ERLC and LHC/FCC based lepton-hadron colliders will essentially enlarge physics search potential of the LHC and FCC for both the SM and BSM phenomena.

Keeping in mind potential of proposed colliders for SM and BSM searches, we invite HEP community to start systematic studies of accelerator, detector and physics search aspects of ERLC and LHC/FCC based *ep* and *eA* colliders.

**Acknowledgement**

The authors are grateful to U. Kaya and B. B. Oner for useful discussions.